# The association between topic growth and citation impact of research publications

Peter Sjögårde[a,b,x], Fereshteh Didegah[b,y]

[a]Health Informatics Centre, Department of Learning, Informatics, Management and Ethics, Karolinska Institutet, Stockholm, Sweden
[b]Metadata for Research Outputs Unit, University library, Karolinska Institutet, Stockholm, Sweden

ORCID:
[x]https://orcid.org/0000-0003-4442-1360
[y]https://orcid.org/0000-0003-0463-0168

Email: peter.sjogarde@ki.se; fereshteh.didegah@ki.se

Corresponding author: Peter Sjögårde, University Library, Karolinska Institutet, 17177 Stockholm, Sweden

## Abstract

Citations are used for research evaluation, and it is therefore important to know which factors influence or associate with citation impact of articles. Several citation factors have been studied in the literature. In this study we propose a new factor, topic growth, that no previous study has studied empirically. The growth rate of topics may influence future citation counts because a high growth in a topic means there are more publications citing previous publications in that topic.

    We construct topics using community detection in a citation network and use a two-part regression model to study the association between topic growth and citation counts in eight broad disciplines. The first part of the model uses quantile regression to estimate the effect of growth ratio on citation counts for publications with more than three citations. The second part of the model uses logistic regression to model the influence of the explanatory variables on the probability of being lowly cited versus being modestly or highly cited. Both models control for three variables that may distort the association between the topic growth and citations: journal impact, number of references, and number of authors.

    The regression model clearly shows that publications in fast-growing topics have a citation advantage compared to publications in slow-growing or declining topics in all of the eight disciplines. Using citation indicators for research evaluation may give incentives for researchers to publish in fast-growing topics, but they may cause research to be less diversified. The results have also some implications for citation normalization.

# 1 Introduction

With the emergence of citation indexing in 1960s, citation indicators became widespread; although they were originally proposed for information retrieval services (Garfield, 1964), soon they were suggested as research evaluation tools by scientometric pioneers (Schubert, Glänzel, & Braun, 1988). They were also used as complementary indicators in research assessment and funding allocation settings such as the UK Research Assessment Exercise or Belgium BOF Key. However, citations are criticized for being one-dimensional and merely measuring the impact of science on the scientific community, and not beyond that. Nonetheless, they are acknowledged as the most popular indicator of scientific impact (Furnham, 1990) and researchers make great efforts to increase the visibility of their research outputs among their peers. Citations have also been found to be associated with quality indicators. For instance, high-quality articles were found to be cited more often (Patterson & Harris, 2009) and articles with more citations were among award winners (Cole, 1973).

Being cited means that the article is seen and used in other research works but the reasons behind citations are yet a matter of debate and vary between researchers and between cited documents (MacRoberts & MacRoberts, 1989). While normative theory of citations holds the view that citations reflect the intrinsic quality of research, social constructivist view casts doubt on this hypothesis as the main motivation for citations and emphasizes on extrinsic factors affecting citation counts of research articles. The latter theory argues that citation behavior may be affected by properties of the article itself (Baldi, 1998; White, 2004). Prior studies have categorized citation factors into groups including properties of the article itself, properties of the journal of the article and properties of the author(s). For example, a great deal of research confirms that articles published in high impact journals tend to be cited more because such journals are perceived to contain high quality content and inherently receive more attention (Haslam et al., 2008). Research collaboration expands authors' networks and increases the chance of higher visibility. As with previous studies, articles with more authors are found to be cited more often than single-author articles and international collaboration is also proposed as an important factor of citation impact. A comprehensive list of references also seems to function as channels to the new published research that increase visibility and impact of articles (Didegah & Thelwall, 2013; van der Pol et al., 2015; Xie et al., 2019).

This study falls in the same vein; it proposes a new factor – topic growth rate – to determine its association with citation impact of articles. Growth of a research field leads to an increase of references to be distributed to previous publications. These references are likely to point to earlier work within the same field but may also be distributed outside the field. This quantitative aspect of growth of research fields may lead to higher average citation rates for early publications within the field, compared to publications of the same age in fields with less growth. The quantitative aspect may not be associated with quality of the publications.

Emergence of a new area is associated with key attributes including novelty, growth, persistence and community (Porter et al., 2019). Growth is gauged through different measures such as number of actors, funders, publications, products, services, etc. and fast growth together with novelty is predicted to lead to prominent impact in the future (Rotolo et al., 2015). Prior studies have already confirmed the association between novelty (newness) and citation impact and showed that new research has a citation advantage (Porter et al., 2019), although it may significantly vary across different subject domains (Thelwall & Sud, 2021).

To the best of our knowledge, no research to date has empirically examined the association between growth rate of publications and citation indicators. This study aims to fill this gap in the literature by addressing the influence of growth rate of research topics on citation counts of their articles and whether it can be a determinant of future citation impact. We control for few important, already well-researched, factors i.e., journal impact factor, number of authors and number of references. In this study, the most granular level of a subject classification that was initially proposed by the first author of current paper is considered as topics.

## 2 Background

Despite the complex nature of citation reasons and motivation, several external factors are known from extensive previous literature as determinants of future citation counts. These factors are properties of the articles and unlike quality, which is subjective, article properties are mostly objective and easy to measure. Nevertheless, research quality is likely a key determinant of citation impact and cannot be replaced by such peripheral factors. For example, an article written by a high impact author and published in a high impact journal may not necessarily be highly cited if it has a low-quality content.

There is a rather large amount of studies addressing the association between citation counts and different factors of articles (for an overview, see Tahamtan et al., 2016). Different categorizations of factors associated with citation rates exists, commonly grouped by properties of the article, journal of publication, references, authors, institutions, countries, and subject fields (Hanssen et al., 2018; Tahamtan et al., 2016; Xie et al., 2019).

Journal citation impact is one of the external factors confirmed to be associated with increased citation counts (Aksnes, 2003; Didegah & Thelwall, 2013; Haslam et al., 2008; Onodera & Yoshikane, 2015). Top journals are perceived to contain higher quality research, so they are cited more; moreover, high impact articles are more likely to get accepted and published in the top journals (Traag, 2021). On the importance of journal impact factor as an indicator of quality, Liu et al. (2015) showed that Ophthalmologic journals with a higher impact factor had a higher peer-reviewed score.

Another important property of articles affecting their future impact is article references. It is a widely held belief that good research is built upon good references. Research works with a higher number of references, especially high impact and recent references, are found to be cited more often (Ahlgren et al., 2018; Mammola et al., 2021). References make the work more visible and may be an indication of the quality of the work. Unusual combinations of cited references (using novel references together with conventional ones) was found to lead to a higher citation impact (Uzzi et al., 2013). Ahlgren et al. (2018) found that reference properties are associated with higher citation impact also when field differences have been normalized.

The growth of references in a field caused by growth in publication outputs may also be a factor causing a higher average citation impact for publications published before or in an early stage of such growth. Vinkler (1996) made a theoretical examination of growth ("scientific development") and citedness. He showed that the publication rate of fields is a determinant factor of citation rates and concluded that "[o]*nly scientific fields or subfields with the same rate of information production offer similar citedness possibilities*". However, this factor has not been addressed empirically.

Research collaboration in terms of number of authors contributing to a paper is also known as a significant determinant of citations in a wide variety of disciplines (Didegah & Thelwall, 2013). If more authors are involved in a study, it is likely that it, in general, results in higher quality research, because more effort and competences have been put into the work. More authors also mean more people who know the authors, get informed of their publications, and possibly cite them in their own works. The wider the collaborative network, the higher the impact and that is probably why international collaborations have a higher influence on citation impact in majority of subject areas (Wagner et al., 2017).

The features of article subject or topic and its association with citation impact of articles have not been widely researched. Such features are the most relevant to the current study. Size of subject fields (in terms of number of publications or number of researchers) has shown to be weakly associated with citation counts in some disciplines (Didegah, 2014). On the other hand, research topic has been found to be important. For instance, the introduction of the h-index by Hirsch in 2005 led to a rapid growth of articles and led to a short term increase in citations in this topic, referred to as a "bubble" by Rousseau et al. (2013). Thelwall and Sud (2021) showed that new research topics are cited more in some disciplines and emerging topics were found to benefit from both within- and outside-field citation links (Kwon et al., 2019).

Bornmann and Tekles (2021) argue that new topics are built upon so called disruptive publications, i.e. publications characterized by their ability to overthrow established thinking (Bornmann et al., 2020). Newness (or novelty) together with the fast growth of publications in an area is predicted to lead to prominent impact in the future (Rotolo et al., 2015) and recent research has shown that new topics have a citation advantage (Thelwall & Sud, 2021). However, the association between the growth of the publication output in a research publication and the citation rates within the field has not been thoroughly investigated. To fill this gap, this article investigates the association between citation impact and topic growth rate in eight broad disciplines, including more than 4,000 topics.

## 3 Data and methods

We used the Karolinska Institutet (KI) in-house version of the Web of Science (WoS) for the analyses, containing data from the Science Citation Index Expanded, Social Sciences Citation Index and Arts & Humanities Citation Index.[1] The KI bibliometric system contains a four level algorithmically constructed publication-level classification that was developed by the first author of this article.

The classification is based on publications from 1995 onwards and registered in the WoS before 2021-03-03, accounting for a total of about 30 million publications and 680 million citation connections. The classification was obtained by clustering the publications in this network using the Leiden algorithm (Traag et al., 2019). Each publication is assigned to exactly one class by the clustering methodology. We normalized citation links using the approach suggested by Waltman and van Eck (2012). Through this approach, each citation link is given the value of one divided by the publication's total number of links.

At the most granular level, the resolution parameter was set in order to obtain classes of about the same size as in Sjögårde and Ahlgren (2018). This was done by testing different values of the resolution parameter, evaluating the size distribution of the outcome and adjusting the resolution until a similar size distribution was obtained. The chosen resolution value resulted in about 118,000 classes after reclassifying classes with less than 50 publications.[2] We refer to the classes at this level as "topics". These topics are areas of research, self-organized through formal communication (citations). Such topics have been contested (Held et al., 2021). However, several studies have shown that classes of this kind correlate with different baselines that have been used for evaluation, such as textual coherence (Boyack et al., 2011; Boyack & Klavans, 2010), article-grant groups (Boyack & Klavans, 2010, 2018), reference lists of "authoritative papers" (Boyack & Klavans, 2018; Klavans & Boyack, 2017; Sjögårde & Ahlgren, 2018), and classifications or controlled vocabulary (Ahlgren et al., 2020; Haunschild et al., 2018). We therefore consider the topics to be at least good enough for the purpose of this study.

Topics were grouped into the parent level using the methodology explained in Sjögårde and Ahlgren (2020), approximating the size of research specialties. At the next level, we used the same methodology to group specialties into research disciplines. The research disciplines were grouped into four major research areas using the same procedure in combination with some manual adjustments.

---

[1] Certain data included herein are derived from the Web of Science ® prepared by Clarivate Analytics ®, Inc. (Clarivate®), Philadelphia, Pennsylvania, USA: © Copyright Clarivate Analytics Group ® 2021. All rights reserved.

[2] We used the console-based version of the Leiden algorithm (version 1.0.0), available at https://github.com/CWTSLeiden/networkanalysis [2020-11-20]. The constant Potts model (CPM) was used, and the resolution parameter was set to 0.000125. We ran 100 iterations (one random start) and allocated 512GB of memory. The process took about 14 hours.

We restricted the analysis to eight disciplines by manually selecting two disciplines from each major research area. We preferred relatively large disciplines and avoided to select closely related disciplines. The disciplines were approximately the same size as classes in classifications commonly used for citation normalization (such as the Web of Science journal categories). The eight disciplines are listed in Table 1 along with the number of publications and the number of topics in each discipline. Labels are presented for the disciplines in Table 1 and throughout the paper. These labels were provided using terms extracted from titles, keywords, journal titles, and author addresses. Terms were defined as a sequence of nouns and adjectives ending with a noun. Term frequency to specificity (TFS) ratio ( Sjögårde et al., 2021) was used to rank the relevance of terms to each class (the alpha value was set to 0.67 at the level of disciplines, giving more weight to frequency than specificity). The three most highly ranked terms were concatenated into a label.

Creating labels for classes in algorithmically constructed classifications is a challenging task. Some terms used to label the disciplines are broader than the focus area of the discipline, for example "medicine" in "medicine; infectious disease; microbiology", and some terms are narrower than the focus area of the discipline, for example "china" in "political science; economics; china". Nevertheless, we believe that the labels are good enough to get a perception of the focus area of the eight disciplines.

The set was restricted to two document types, i.e. "article" and "review". The sizes of topics were too small to consider reviews individually, we therefore choose to treat the two types combined. Conference papers were omitted because of their sparse and even more highly skewed citation distribution. Furthermore, the analyses were restricted to the publication year 2015 and citations were counted from 2015 to March 2021, which gives a citation window of at least five full years. Hence, we use a "diachronous" or "prospective" perspective on citations (Glänzel, 2004).

Growth rates in small topics are unstable and publications in such topics were excluded. Topics were regarded as too small if containing less or equal to 15 publications in 2013-2015 or 2016-2018 (see section 3.2.1 for further detail). The column "# Publ. (included)" in Table 1 presents the number of publications in each discipline after discarding small topics.

Table 1: Discipline labels, number of publications and number of topics within each discipline. "# Publ. (total)" gives the number of publications in each class in 2015 and "# Publ. (included)" the number of publications after excluding publications in small topics. "# Topics" gives the number of topics after excluding small topics".

| Discipline (machine generated label) | # Publ. (total) | # Publ. (included) | # Topics |
|---|---|---|---|
| psychology; cognition; cognitive neuroscience | 15,869 | 14,147 | 676 |
| medicine; infectious disease; microbiology | 8,925 | 7,888 | 347 |
| chemistry; graphene; chemical engineering | 20,029 | 18,356 | 628 |
| computational science; electronics; computational engineering | 13,464 | 10,749 | 512 |
| education; higher education; teaching | 11,375 | 8,825 | 482 |
| political science; economics; china | 10,272 | 8,074 | 486 |
| plant science; plant physiology; agriculture | 15,252 | 12,961 | 688 |
| geology; earth science; earth | 14,987 | 13,349 | 714 |

## 3.1. Dependent variable

We used citation counts as the outcome variable. We did not use normalization because each discipline was analysed separately and the analysis was restricted to only one publication year. The distributions of citation counts are highly skewed in all eight disciplines, as shown in Figure 1. The y-axes in Figure 1, the citation count variable, is in a log-10 scale.

## 3.2. Explanatory variables

### 3.2.1. Growth ratio

We are primarily interested in the effect of topic growth on citation counts and topics are defined as classes at the most granular level in the classification. We use the formula proposed by Wang (2018) to calculate the growth ratio for topic $i$ in year $t$:

$$r_{i,t,\Delta t} = \frac{p_{i,t,\Delta t}}{p_{i,t}} \quad (1)$$

where $p_{i,t}$ is the number of publications for topic $i$ in year $t$.

$\Delta t$ is a time interval and $p_{i,t,\Delta t}$ is the number of publications in year $t + \Delta t$.

We are interested in explaining the citation counts for publications published in 2015 by its growth after the time of publication. Therefore, $t$ was set to 2015, and $t + \Delta t$ to 2018. Wang (2018) points out that publication counts may fluctuate randomly between years. To smoothen out such random fluctuation, we used a three-year average of the publication count (as proposed by Wang, 2018):

$$r_{i,t,\Delta t} = \frac{\bar{p}_{i,t,\Delta t}}{\bar{p}_{i,t}} \quad (2)$$

where $\bar{p}_{i,t}$ is the average number of publications for topic $i$ in year $t$, $t$-1 and $t$-2 and $\bar{p}_{i,t,\Delta t}$ is the average number of publications for topic $i$ in year $t + \Delta t$, $t + \Delta t$ -1 and $t + \Delta t$ -2.

In summary, growth rate was calculated as the ratio between the average number of publications in a topic in the time period 2016-2018 and three years earlier, i.e. 2013-2015.

To further avoid fluctuations caused by small number of publications in the topics, we restricted the publication set to topics with $\bar{p}_{i,t} > 5$ and $\bar{p}_{i,t,\Delta t} > 5$.

### 3.2.2. Control variables

We controlled for three variables, all known to be positively associated with citation counts in the previous literature: (1) Journal impact factor (JIF), (2) Number of authors and (3) Number of references. We used the JIF from the year of the publications, i.e. 2015. Histograms for all variables are presented in Figure 1 and the average of each variable is presented in Table 2. Scatter plots of the association between citation counts and each of the four explanatory variables are illustrated in Figure 2.

*Table 2: Averages of the variable values for each of the eight disciplines. The averages are calculated per publication across all topics per discipline.*

| Discipline (machine generated label) | Avg. no. of citations | Avg. no. of authors | Avg. no. of references | Avg. growth ratio | Avg. JIF |
|---|---|---|---|---|---|
| psychology; cognition; cognitive neuroscience | 19.4 | 4.1 | 54.7 | 1.1 | 3.2 |
| medicine; infectious disease; microbiology | 17.2 | 6.5 | 35.5 | 1.2 | 3 |
| chemistry; graphene; chemical engineering | 46.2 | 5.5 | 44.4 | 1.6 | 5.3 |
| computational science; electronics; computational engineering | 12 | 3.7 | 31.6 | 1.4 | 1.6 |
| education; higher education; teaching | 10.9 | 3 | 48.7 | 1.2 | 1.5 |
| political science; economics; china | 9.9 | 1.9 | 50.1 | 1.2 | 1.2 |
| plant science; plant physiology; agriculture | 18.8 | 5.6 | 50.6 | 1.1 | 2.9 |
| geology; earth science; earth | 17.6 | 4.6 | 65.9 | 1.2 | 2.8 |

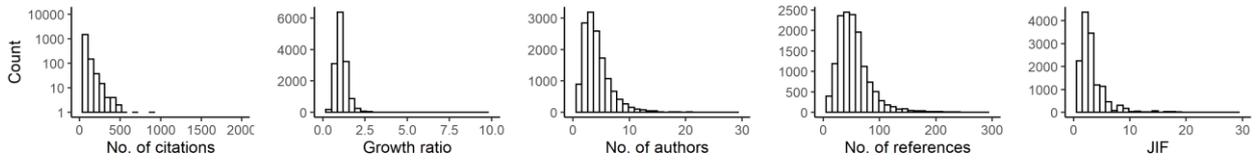
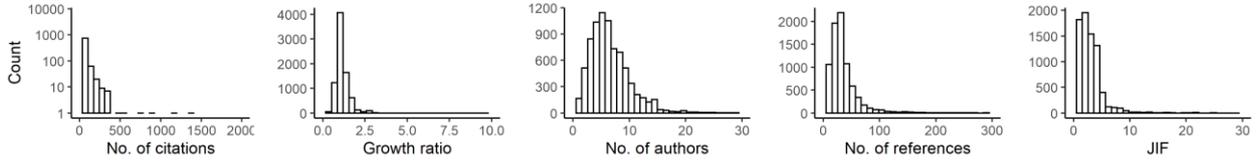
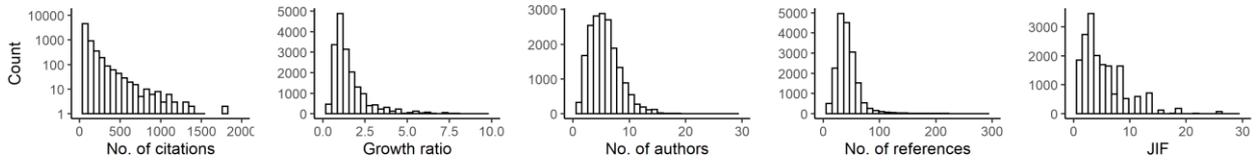
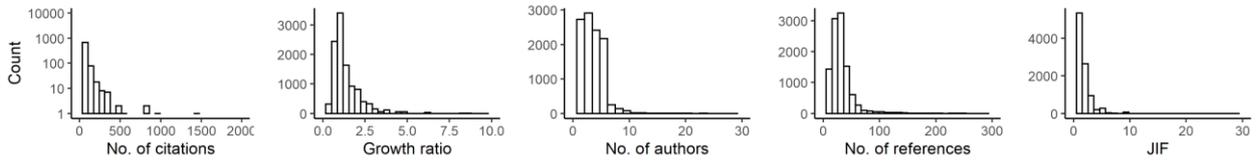
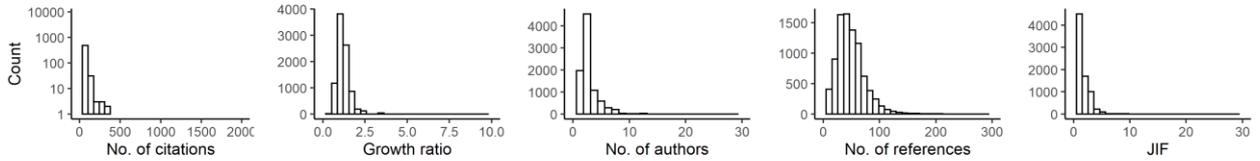
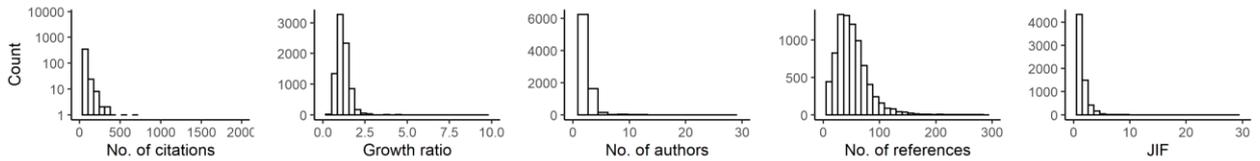
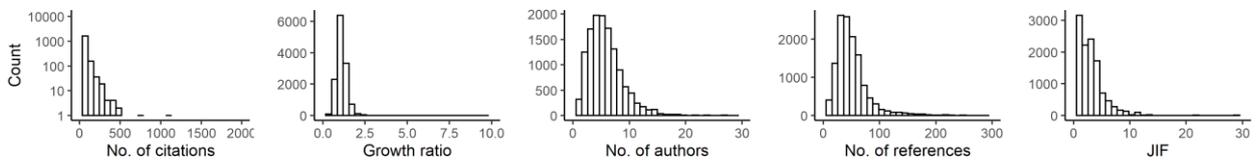
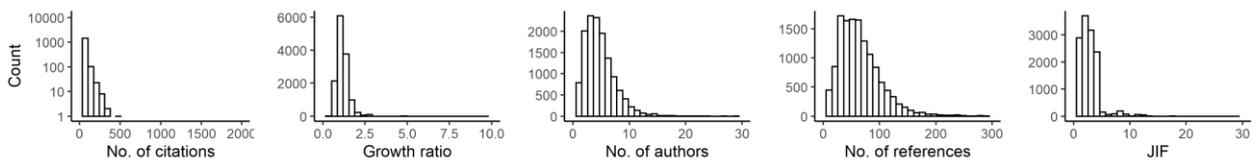

*Figure 1: Histograms of the 5 variables for each of the eight disciplines. For citation counts a log-10 scale is used on the y-axis. To improve readability of the histograms, the x-axis has been restricted to 2,000 citations, a growth-ratio of 10, 30 authors, 300 references and a JIF value of 30. Some outliers are omitted by these restrictions. Outliers are included in the scatter plots in Figure 2.*

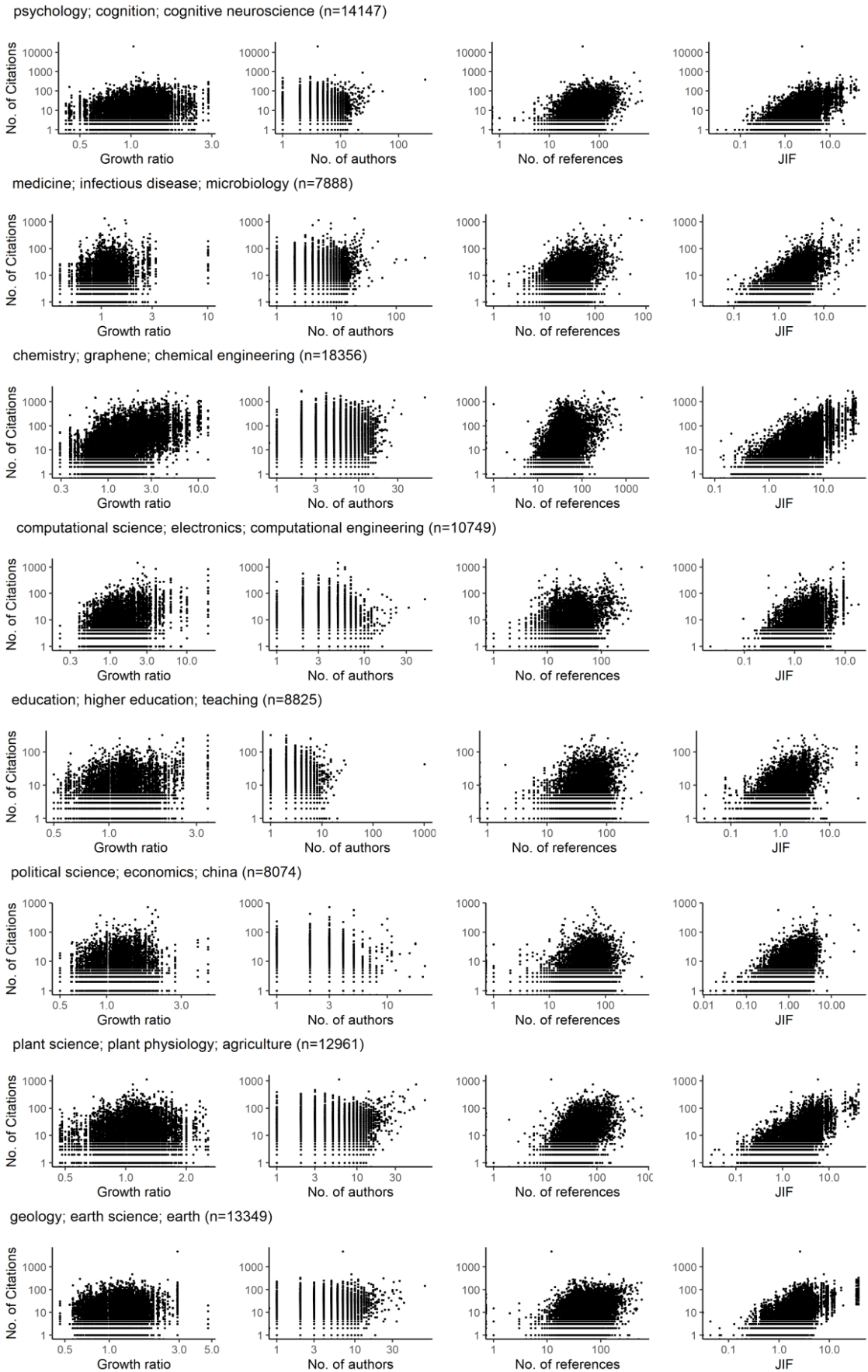

*Figure 2: Scatter plots of the number of citations and each of the four explanatory variables and each of the eight disciplines. Log-10 scale used on both axis.*

## 3.3. Statistical procedures

Citation counts are highly skewed with an excessive number of publications with no or just a few citations, as shown in Figure 1. Regression models that assume a normal distribution of the dependent variable are not appropriate. Quantile regressions are commonly used in scientometrics because of the skewed distributions. However, quantile regression assumes continuous data. The high number of publications with no or few citations may lead to biased or erroneous results for lower quantiles. This problem has been addressed by Shahmandi et al. (2021), proposing a two-part hurdle quantile regression model. They propose a hurdle at three, based on common citation distributions. We use this model to estimate the effect of the explanatory variables on citation counts. The model groups the observations in two parts, one including publications with more than three citations and one including publications with less or equal to three citations. The first part of the model uses quantile regression to estimate the effect of the explanatory variables on citation counts for publications with more than three citations. The second part of the model uses logistic regression to model the influence of the explanatory variables on the probability of being cited more than three times versus 0-3 times.

For the quantile regression, we used the R-function *bayesQR()* from the package with the same name. We estimated the regression for deciles 0.1 to 0.9. 10,000 iterations were used (the *ndraw* parameter) and we kept every $10^{th}$ draw (*keep*). The function did not converge for a couple of quantiles, these are omitted in the results. We skipped the first 500 of the 1,000 kept results for the summary (*burnin* parameter in the *summary()* function).

For the logistic regression, we used the R-function *glm()* from the *stats* package to model the probability of low citation counts (<=3) versus modest or high citation counts (>3). A binomial model was used (*binomial(link="logit")*). The estimate from the logistic regression corresponds to the log of the odds ratio. We therefore present the exponent of the estimate to facilitate easier interpretations of the results.

## 4 Results

## 4.1. Quantile regression

Figure 3 shows the effect on citations for each of the explanatory variables when increasing the explanatory variable with one unit. Bayes estimates are given at each of the considered quantile levels using a credible interval between 2.5% and 97.5%. The growth ratio of the topic has a positive influence on citation counts in all eight disciplines and at all estimated quantile levels. The effect increases steadily when moving from low to high quantiles. The largest effect can be seen in "chemistry; graphene; chemical engineering" in which an increase of one unit is associated with about 30 more citations at quantile 0.9. This is not surprising, considering that "chemistry; graphene; chemical engineering" has the highest citation rates. In the middle of the distribution (quantile 0.5), the estimate ranges from an increase of about 2.5 to about 13 citations per unit growth ratio in the same discipline.

The number of references and JIF has a positive effect on the number of citations. An increasing association at higher quantiles is also observed for these variables. The effect size for the number of references ranges from about 0.026 in "political science; economics; china" to 0.123 in "medicine; infectious disease; microbiology" per unit increase at quantile 0.5. It should be noted that the scales of the variables differ substantially. As an example, the average number of references in "medicine; infectious disease; microbiology" is about 36. An increase of 36 references is associated with 18 more citations at quantile 0.5 in this discipline.

The bayes estimate for the JIF is around three in most disciplines. For example, in "geology; earth science; earth" an increase of one unit of the JIF is associated with an increase of about 2.8 citations at quantile 0.5 and about 6.4 citations at quantile 0.9. The average JIF in the discipline is about 2.8. An increased JIF of 2.8 is associated with about 18 more citations.

The association between the number of authors and the number of citations follow approximately the same pattern as the other explanatory variables, but with some exceptions. In "education; higher education; teaching", the effect size does not increase at higher quantiles. In "psychology; cognition; cognitive neuroscience", there is a dip for the highest quantile (0.9) and in "chemistry; graphene; chemical engineering", there is a dip for the two highest quantiles (0.8 and 0.9). The credible intervals are generally larger for the number of authors.

## 4.2. Logistic regression

The logistic regression shows a strong association between growth ratio and low citation counts ($<=3$) versus moderate to high citation counts ($>3$). The odds ratio varies from 1.68 in "political science; economics; china" to 3.0 in "psychology; cognition; cognitive neuroscience". This shows that it is 1.68 times more likely in the former discipline that a publication belongs to the group with more than three citations if the growth ratio increases with one unit; in the latter discipline it is three times more likely. In all the eight disciplines, it is substantially more likely to be cited more than three times if the publication belongs to a fast-growing research topic.

The control variables also show significant positive associations with citation counts. Adding an author is associated with an increase of 0.09 (in "computational science; electronics; computational engineering") to 0.2 (in "political science; economics; china") of the odds of membership in the group with more than three citations. An increase of one reference is associated with an increase of 0.014 (in "political science; economics; china") to 0.029 (in "medicine; infectious disease; microbiology") of the odds of membership in the group with more than three citations. And for each unit of increase in the JIF, the change of odds of membership in the group with more than three citations increases with 0.92 (in "psychology; cognition; cognitive neuroscience") to 3.32 (in "political science; economics; china").

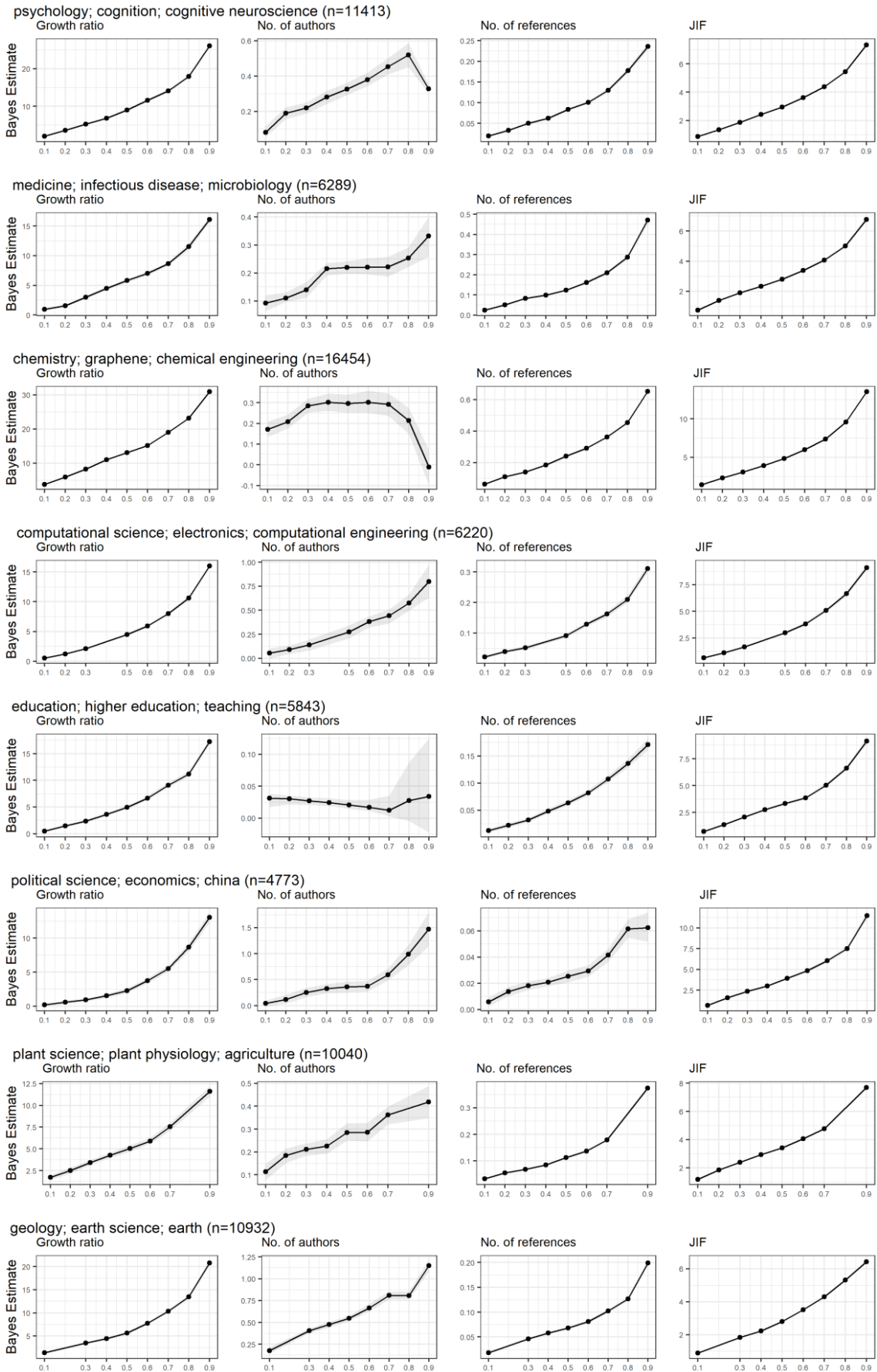

*Figure 3: Bayes estimates of the quantile regression model for each of the eight disciplines and the four explanatory variables.*

*Table 3: Results of the logistic regression model for each of the eight disciplines and the four explanatory variables.*

**psychology; cognition; cognitive neuroscience (n=14147)**

|  | Estimate | Exp. Estimate | SE | z-value | p-value | Signif. |
|---|---|---|---|---|---|---|
| (Intercept) | -2.6 | 0.07427 | 0.12 | -22 | 9.8E-111 | *** |
| growth_ratio | 1.1 | 3.004 | 0.08 | 14 | 1.6E-44 | *** |
| num_authors | 0.091 | 1.095 | 0.012 | 7.6 | 3.2E-14 | *** |
| num_references | 0.019 | 1.019 | 0.0011 | 18 | 1.2E-68 | *** |
| jif | 0.65 | 1.916 | 0.023 | 28 | 1.4E-177 | *** |

**medicine; infectious disease; microbiology (n=7888)**

|  | Estimate | Exp. Estimate | SE | z-value | p-value | Signif. |
|---|---|---|---|---|---|---|
| (Intercept) | -2.4 | 0.09072 | 0.15 | -16 | 4.5E-57 | *** |
| growth_ratio | 0.63 | 1.878 | 0.099 | 6.4 | 2.1E-10 | *** |
| num_authors | 0.098 | 1.103 | 0.011 | 8.5 | 1.2E-17 | *** |
| num_references | 0.029 | 1.029 | 0.0022 | 13 | 6.2E-40 | *** |
| jif | 0.75 | 2.117 | 0.031 | 24 | 2.4E-132 | *** |

**chemistry; graphene; chemical engineering (n=18356)**

|  | Estimate | Exp. Estimate | SE | z-value | p-value | Signif. |
|---|---|---|---|---|---|---|
| (Intercept) | -2.6 | 0.07427 | 0.12 | -23 | 6.7E-114 | *** |
| growth_ratio | 0.88 | 2.411 | 0.058 | 15 | 7.7E-53 | *** |
| num_authors | 0.1 | 1.105 | 0.013 | 7.6 | 4E-14 | *** |
| num_references | 0.024 | 1.024 | 0.0019 | 13 | 2.5E-38 | *** |
| jif | 0.77 | 2.16 | 0.024 | 32 | 4.8E-230 | *** |

**computational science; electronics; computational engineering (n=10749)**

|  | Estimate | Exp. Estimate | SE | z-value | p-value | Signif. |
|---|---|---|---|---|---|---|
| (Intercept) | -2.9 | 0.05502 | 0.091 | -32 | 5.3E-225 | *** |
| growth_ratio | 0.62 | 1.859 | 0.036 | 17 | 4.4E-68 | *** |
| num_authors | 0.086 | 1.09 | 0.014 | 6.3 | 3.6E-10 | *** |
| num_references | 0.024 | 1.024 | 0.0016 | 15 | 3.8E-52 | *** |
| jif | 1 | 2.718 | 0.03 | 33 | 1.8E-237 | *** |

**education; higher education; teaching (n=8825)**

|  | Estimate | Exp. Estimate | SE | z-value | p-value | Signif. |
|---|---|---|---|---|---|---|
| (Intercept) | -2.8 | 0.06081 | 0.12 | -23 | 3.2E-117 | *** |
| growth_ratio | 0.86 | 2.363 | 0.076 | 11 | 3.9E-30 | *** |
| num_authors | 0.11 | 1.116 | 0.016 | 7.2 | 7.2E-13 | *** |
| num_references | 0.022 | 1.022 | 0.0012 | 18 | 2.1E-72 | *** |
| jif | 0.91 | 2.484 | 0.036 | 25 | 6.6E-140 | *** |

**political science; economics; china (n=8074)**

|               | Estimate | Exp. Estimate | SE    | z-value | p-value  | Signif. |
|---------------|----------|---------------|-------|---------|----------|---------|
| (Intercept)   | -2.6     | 0.07427       | 0.12  | -22     | 1.3E-109 | ***     |
| growth_ratio  | 0.52     | 1.682         | 0.072 | 7.3     | 4.1E-13  | ***     |
| num_authors   | 0.18     | 1.197         | 0.025 | 7.3     | 2.5E-13  | ***     |
| num_references| 0.014    | 1.014         | 0.001 | 14      | 5E-43    | ***     |
| jif           | 1.2      | 3.32          | 0.043 | 29      | 2.9E-181 | ***     |

**plant science; plant physiology; agriculture (n=12961)**

|               | Estimate | Exp. Estimate | SE    | z-value | p-value   | Signif. |
|---------------|----------|---------------|-------|---------|-----------|---------|
| (Intercept)   | -3       | 0.04979       | 0.13  | -23     | 5.80E-118 | ***     |
| growth_ratio  | 0.96     | 2.612         | 0.09  | 11      | 1.20E-26  | ***     |
| num_authors   | 0.081    | 1.084         | 0.011 | 7.2     | 5.80E-13  | ***     |
| num_references| 0.024    | 1.024         | 0.0016| 16      | 4.50E-55  | ***     |
| jif           | 0.95     | 2.586         | 0.028 | 35      | 2.20E-262 | ***     |

**geology; earth science; earth (n=13349)**

|               | Estimate | Exp. Estimate | SE      | z-value | p-value  | Signif. |
|---------------|----------|---------------|---------|---------|----------|---------|
| (Intercept)   | -2.9     | 0.05502       | 0.13    | -23     | 5.1E-116 | ***     |
| growth_ratio  | 0.76     | 2.138         | 0.08    | 9.5     | 1.9E-21  | ***     |
| num_authors   | 0.15     | 1.162         | 0.012   | 12      | 6.9E-33  | ***     |
| num_references| 0.017    | 1.017         | 0.00095 | 18      | 3.3E-71  | ***     |
| jif           | 0.94     | 2.56          | 0.027   | 34      | 2.5E-260 | ***     |

Note: Signif. codes:  0 '***' 0.001 '**' 0.01 '*' 0.05 '.' 0.1 ' ' 1

## 5 Discussion and conclusion

We have investigated the impact of topic growth on citation impact of articles in eight different disciplines. To ensure that the impact of topic growth on citations is not distorted by other factors, three important factors, i.e. journal impact, number of authors and number of references, were controlled for any interference with the citation impact association. The results of regression models for these factors in the current study also clearly confirm their importance for the citation impact of articles. However, other factors such as author prestige and impact in the field or open access status of the articles could have been influential. A more comprehensive model, controlling for more factors, could help further reduce the risk of distorting the picture of the association between the topic growth and citations counts to articles.

The findings confirm that articles published in the topics growing after the time point of the publication received more citations in general in all of the eight disciplines examined. The results from the two-part regression model shows that articles in growing topics are more likely to be at least modestly cited ($> 3$) and that the number of citations is associated with the growth of the topic after publication. Even though we have chosen broad disciplines to get a wide coverage over research fields, these results cannot be generalized to the entire science; there might be some disciplines that do not follow the patterns observed in the eight disciplines and future research may focus on other fields.

In this study, the association between citation impact and growth of topics is investigated at a quantitative level and what causes the relations has not been scrutinized. In particular, the relation between increased citation impact of papers in growing topics and research quality is not known. Moreover, the topics used in the study have been created using community detection in citation networks. Little is known about temporality of such topics, for example this methodology's ability to capture the formation of new topics. The operationalization of topics may affect the results. Future studies may explore how new topics emerge and how these dynamics are captured by community detection in citation networks or other similar classification approaches.

According to Small et al. (2014), growth together with newness are the main properties of emerging topics, and the two factors are predicted to contribute to prominent impact in the future (Rotolo et al., 2015). Due to its intrinsic newness, a novel topic is not completely explored and most aspects of it may have yet remained unknown. To become a hot emerging research topic, it also needs to grow in volume. The novelty together with the volume of publications attract other researchers' attention and increase visibility and impact (Tu & Seng, 2012). We have not studied novelty in this paper but Thelwall and Sud (2021) found that articles published around novel topics have a citation advantage using a word frequency analysis. It would be interesting to study topic novelty and growth simultaneously for their citation advantage in a future study.

At a publication level, new or disruptive publications can be detected through bibliographic coupling and uncoupling and refer to "something new that eclipses attention to previous work upon which it has built" (Wu et al., 2019). The growing number of publications in a topic may be due to disruptive publications existing in the topic (Funk & Owen-Smith, 2017; Wu et al., 2019). Bornmann and Tekles (2021) argue that indicators of disruption is linked to the theory of Kuhn (1996) in which science is not seen as stable and cumulative, but rather as occasionally shifting focus and being transformed by novel ideas and breakthroughs. Disruptive publications may therefore be the starting point of new topics and may spur a growing number of publications in the topic, and thereby a lot of citations to the early publications within the topic. Consequently, disruptive publications are likely to be highly cited and be found in topics with a high growth rate after being published. For disruptive publications, it can be argued that a high citation impact reflects quality of the publications, because they have advanced research by defining new topics. However, only a small number of publications published in an early phase of an emerging topic can be expected to be disruptive. There are likely publications adapting the new topic in an early stage that are more mediocre regarding innovativeness and quality. Such publications may still have a citation advantage, because the topic is growing and so is the number of publications citing previous research within the topic. For such publications, it can be argued that the increased citation rates are not an indication of quality, but merely associated with the early adaption.

The growth of a topic may be the result of a hype that leverages the publication output in the topic as well as citations to (early) works in the topic (Rousseau et al., 2013). Researchers may benefit, in terms of citations and attention, from publishing in such topics. This can be seen as a topic-level Matthew effect where already recognized (hyped) topics are addressed by researchers, causing articles in the topic to be more cited (Biglu, 2007; Larivière & Gingras, 2010; Merton, 1968). The use of citation indicators for research evaluation may increase the incentives for researchers to publish in hyped topics. This may be an unwanted effect and it may cause research to be less diverse. Responsible use of citation metrics should take such effects into account and recognize the complexity of reasons behind citations and that higher number of citations may reflect other factors than quality and impact (Hicks et al., 2015).

The association between topic growth and citations counts has implications for citation normalization. In this study we have compared topics within eight disciplines and restricted the analysis to one publication year. By doing so, we restrict each analysis to a category that can be used for classification-based (cited-side) normalization. Therefore, we expect that publications in fast-growing topics generally have a higher normalized citation rate than publications in slow-growing or declining topics, when classification-based normalization is used at this level of granularity. If the normalization is based on more granular categories, the effect of topic growth on normalized citation impact is likely to be reduced. However, too granular categories have disadvantages (Ruiz-Castillo & Waltman, 2015; Waltman & van Eck, 2013). The smaller the categories, the more unstable are the statistical properties. Furthermore, using small normalization categories may reduce the calculated citation impact between articles in fields with different "real" impact and booze the citation impact of articles in fields with reference values that are arbitrarily low. Also item-oriented normalization procedures (Colliander, 2015; Colliander & Ahlgren, 2019) can be expected to reduce the effect of topic growth on the calculated citation impact. Item-oriented normalization typically makes use of small reference sets and may suffer from the same disadvantages as finely granular clustering-based approaches. The effect of the association between topic growth and citation counts on citing-side normalization (Leydesdorff & Opthof, 2010; Li & Ruiz-Castillo, 2013; Waltman et al., 2013) is less intuitive and remains unclear. We suggest that the association between topic growth and citation counts should be taken into account when citation normalization procedures are developed and evaluated and that future work on this topic investigates the relation between different normalization procedures and topic growth.

## Declarations

**Acknowledgments**
We would like to thank Per Ahlgren and two anonymous reviewers for their relevant and constructive comments on an earlier version of this paper.

**Author contributions**
Peter Sjögårde: Conceptualization, Methodology, Formal analysis, Writing - Original Draft, Writing - Review & Editing, Visualization
Fereshteh Didegah: Conceptualization, Methodology, Writing - Original Draft, Writing - Review & Editing

**Funding**
Peter Sjögårde was funded by The Foundation for Promotion and Development of Research at Karolinska Institutet.

**Availability of data and material**
Data are subject to copyright by Clarivate Analytics ®.

**Code availability**
R code used for the analyses is available upon request.

**Conflict of interest**
The authors declare no competing financial interests.